\documentclass[a4paper,12pt]{article}

\usepackage{amsmath}\usepackage{amssymb}
\usepackage{latexsym}\usepackage{cite}

\newcommand{\be}{\begin{equation}}
\newcommand{\ee}{\end{equation}}

\def\dspace{\baselineskip = .25in}
\def\beq{\begin{equation}}
\def\eeq{\end{equation}}

\def\al{\alpha}
\def\bt{\beta}

\def\ga{\gamma}
\def\de{\delta}
\def\De{\Delta}

\def\te{\theta}

\def\lam{\lambda}

\def\ep{\epsilon}

\def\sq{\sqrt}

\def\l{\left (}
\def\r{\right )}

\def\fr{\frac}
\def\la{\label}
\def\hs{\hspace}
\def\vs{\vspace}

\def\ran{\rangle}
\def\lan{\langle}
\def\ov{\overline}
\def\tl{\tilde}
\def\tm{\times}

\begin{document}

\begin{flushright}
BA-06-17\\
OSU-HEP-06-11\\
\end{flushright}

\vs{0.5cm}

\begin{center}
{\Large\bf

$\te_{13}$, Rare Decays and Leptonic CP Violation
}
\end{center}

\vspace{0.5cm}
\begin{center}
{\large
{}~Qaisar Shafi$^{a}$~~and~
{}~Zurab Tavartkiladze$^{b}$
}
\vspace{0.5cm}

$^a${\em Bartol Research Institute, Department of Physics and Astronomy

University of Delaware, Newark, DE 19716, USA \\

$^b$Department of Physics, Oklahoma State University, Stillwater, OK 74078, USA
}

\end{center}
\vspace{0.6cm}

\begin{abstract}

\dspace

We introduce a flavor ${\cal U}(1)$ symmetry \`a la Froggat-Nielsen to explain the
magnitude of the CKM mixing angles as well bilarge mixing
in the neutrino sector. In the framework of GUTs we also estimate the magnitude of the
third leptonic mixing angle $\te_{13}$. Estimates for rare leptonic decays
such as $\mu \to e \ga $ are also provided. Finally, leptogenesis is
exploited for estimating the magnitude of leptonic CP violation.

\end{abstract}

\vs{0.7cm}

\dspace

\newpage

\section{Introduction: Models of Neutrino Oscillations and\\
${\cal U}(1)$ Flavor Symmetry}

In recent years atmospheric \cite{Fukuda:2000np} and solar
\cite{Fukuda:2001nj} neutrino data have confirmed the existence of neutrino
oscillations. A global analysis \cite{Maltoni:2004ei},
\cite{Fogli:2005gs} for the oscillation
parameters yields
$$
\De m_{\rm sol}^2=7.9\cdot 10^{-5}{\rm eV}^2~,~~~~ \De m_{\rm
atm}^2=2.4\cdot 10^{-3}{\rm eV}^2~,
$$
 \beq
 \sin^2\te_{12}=0.314~,~~~~\sin^2\te_{23}=0.44~.
  \la{data}
  \eeq
Apart from the fact that at least two of the neutrinos are
massive, the most
exciting aspect about this data is the fact that it indisputably
points to new physics beyond the SM/MSSM\footnote{Later we will consider
supersymmetric constructions because of various theoretical and phenomenological
motivations. However, the discussion concerning  the fermion sector
can be applied  also to non-SUSY constructions.}.
This is so because the  scale corresponding
to the atmospheric anomaly is $\approx 0.05$~eV, while at the renormalizable
level, both the SM and MSSM (with no matter parity violation, i.e. stable LSP)
gives $m_{\nu }  = 0$. Non-renormalizable
$d=5$ operators $\fr{(lh_u)^2}{M_*}$ contribute to the neutrino mass, and
with a cut off $M_*\sim M_{\rm Pl}=2.4\cdot 10^{18}$~GeV
give  rise to neutrino masses
$\de m_{\nu } \sim (10^{-5}-10^{-4})$~eV, which is way too small. Therefore,
some extension of the SM/MSSM is needed, and presumably the simplest way is to introduce right-handed
neutrino (RHN) states $\nu^{c}$.
With suitably arranged Dirac and Majorana couplings of
the $\nu^c$s,  neutrino masses of the desired magnitude can be
generated via the  see-saw mechanism \cite{seesaw}.

The third leptonic mixing angle
$\te_{13}$ has not yet been measured  but a useful upper bound exists,
\beq
\te_{13}\stackrel{<}{_\sim }0.2~,
  \la{chooz}
  \eeq
provided by the
CHOOZ experiment \cite{Apollonio:2002gd}.
Looking at (\ref{data}) and (\ref{chooz}) one can
try to understand the origin of two large mixing angles
($\te_{12}\simeq 35^{\rm o}$, $\te_{23}\simeq 42^{\rm o}$) and a
suppressed third angle ($\te_{13}\stackrel{<}{_\sim }12^{\rm o}$).
It is desirable to have a framework which, together with neutrino masses,
provides some understanding of these facts. One should note that
 the lepton mixing matrix $U^l_{\al i}$ (where $\al , i=1,2,3$ are indices
which label
the flavor and neutrino mass eigenstates respectively), carries contributions not only from the
neutral lepton sector  but also from the charged lepton sector.
An understanding of lepton mixings requires a close look also at the charged fermion sector.

The hierarchies between charged fermion Yukawa couplings and the CKM
mixing elements can be parameterized by the parameter $\lam \simeq 0.2$
as follows:
$$
\lam_t\sim 1~,~~~~~\lam_u:\lam_c:\lam_t\sim \lam^8:\lam^4:1~.
$$
$$
\lam_b\sim \lam_{\tau }\sim \fr{m_b}{m_t}\tan \bt ~,~~~~
\lam_d:\lam_s:\lam_b\sim \lam^4:\lam^2:1~.
$$
$$
\lam_e:\lam_{\mu }:\lam_{\tau }\sim \lam^5:\lam^2:1~.
$$
\beq
V_{us}\approx \lam ~,~~~~V_{cb}\approx \lam^2 ~,~~~~V_{ub}=\lam^4-\lam^3 ~,
\la{obsmass}
\eeq
where the MSSM parameter $\tan \bt =\fr{v_u}{v_d}$.
The lepton mixing angles are large in comparison to the CKM mixing elements,
with $\te_{13}$  a possible exception. What is the reason for this apparent mismatch between the lepton and quark
mixings?

Since symmetries play a central role in particle physics, our intention is to
gain some understanding of the observed mass hierarchies and the magnitudes of the various  mixing angles using
symmetry arguments. We can postulate a  flavor symmetry $G_F$ which acts in the
flavor space and require invariance of the action under the transformation
\beq
f_{\al }\to (U_F)_{\al }^{\bt }f_{\bt }~,
\la{fltr}
\eeq
where $f_{\al}$ indicates the fermion flavor and $U_F $ is a transformation matrix
(in flavor space) determined by the appropriate  representation of $G_F$.
Since the Lagrangian is restricted by $G_F$,
the structure of the fermion mass matrices are (completely) determined by the
flavor symmetry breaking pattern.

The simplest realization of this picture is to consider an Abelian ${\cal U}(1)$
flavor symmetry \cite{Froggatt:1978nt}. Then the transformation in (\ref{fltr}) is
$ (U_F)_{\al }^{\bt }=e^{iQ[f_\al]}\de_{\al }^{\bt }$, where $Q[f_{\al }]$
is the ${\cal U}(1)$ charge of $f_{\al }$.
By an appropriate charge assignment it is possible to generate the desired
 hierarchies between charged fermion masses and the CKM mixing elements. However,
a self consistent neutrino sector needs a more careful treatment.
First of all, one should generate neutrino masses $m_i$ with the property
\beq
\fr{|m_2^2-m_1^2|}{|m_3^2-m_2^2|}\simeq \fr{1}{30}\sim \lam^2~.
\la{2nmasrat}
\eeq
It is desirable to have an understanding of this `little' hierarchy based on
${\cal U}(1)$. Moreover, the lepton mixing angles compatible with
(\ref{data}), (\ref{chooz}) should be generated.
The task becomes more involved if we wish to realize this program within Grand Unification (GUT), which
relates quarks and leptons. Before discussing GUTs, we consider the
possibilities offered by ${\cal U}(1)$ flavor symmetry to realize
 bi-large neutrino mixings and suppressed $\te_{13}$.

The data in (\ref{data}) says nothing, in principle, about absolute values of neutrino
masses\footnote{The constraint on the sum of the absolute values of neutrino masses provided by combining the recent WMAP  \cite{WMAP} and
galaxy clustering data   is $\sum_i|m_i|\leq 0.68$~eV. From the WMAP three-year data alone the bound found in \cite{Fukugita:2006rm}
is $\sum_i|m_i|\leq 2.0$~eV.}.
This leaves us with several
distinct possibilities for the neutrino mass pattern.

{\bf 1.} Three (quasi)-degenerate neutrinos.

We may have a situation with
\beq
m_1\simeq m_2\approx m_3\gg \sq{\De m_{\rm atm}^2}~,
\la{deg}
\eeq
in which case tiny splittings between masses are required  to explain neutrino oscillations. This pattern
has some theoretical problems. Neutrino mass matrices
with `near' degenerate mass spectrum usually suffer from renormalization group (RG) instabilities
\cite{Ellis:1999my}.
The RG corrections may either give undesirable splitting between masses or they may
fix the mixing angles to be small.
In addition, it is difficult to generate degenerate neutrino mass texture
with the help of only ${\cal U}(1)$ flavor symmetry. For this purpose
non-Abelian symmetries are usually more efficient \cite{Wetterich:1998vh}. With local non-Abelian
symmetries,  special care must be exercised to ensure anomaly
cancellations.

{\bf 2.} Inverted hierarchical neutrinos.

In this case the neutrino spectrum is
\beq
m_1\simeq m_2~,~~~m_3\approx 0~.
\la{inv}
\eeq
This spectrum, when confronted with the data in (\ref{data}) dictates that
$m_{1,2}\simeq \sq{\De m_{\rm atm}^2}$,
$m_3\stackrel{<}{_\sim }\sq{\De m_{\rm sol}^2}$. It is possible to achieve this spectrum via couplings respecting
$L_e-L_{\mu }-L_{\tau }$
symmetry \cite{Petcov:1982ya}, or by ${\cal U}(1)$ flavor symmetry
\cite{Shafi:2000su}, \cite{Shafi:2001se}, \cite{Leontaris:2004rd}. However,
the angle $\te_{12}$, it turns out is nearly maximal which is excluded.
The texture is stable under RG corrections
\cite{Shafi:2001se},
and no improvement from radiative effects is expected. To be viable, there should be
contributions which strongly violate the ${\cal U}(1)$ symmetry
\cite{Leontaris:2004rd}.
However, this must be done in such a way as to maintain the hierarchy in
(\ref{2nmasrat}), keeping at the same time $\te_{13}$ adequately suppressed.
This requires a certain amount of fine tuning.

{\bf 3.} Hierarchical neutrinos.

This scenario has the following mass pattern
\beq
m_1\stackrel{<}{_\sim } m_2\ll m_3~,
\la{hier}
\eeq
which means that $m_3\simeq \sq{\De m_{\rm atm}^2}$ and
$m_2\sim \sq{\De m_{\rm sol}^2}$. One should note that there is a wide class
of neutrino textures which provide this spectrum. It includes
democratic textures \cite{Tanimoto:1999pj} which are compatible
with  ${\cal U}(1)$
flavor symmetry \cite{Shafi:2002qh}. The (normal) hierarchical structures
are usually stable under RG effects \cite{Antusch:2003kp}, and for these reasons will be the main focus of our attention
for the remainder of the paper.
We would like to stress, however, that the other two scenarios are not excluded
by the experimental data.

The democratic approach \cite{Tanimoto:1999pj}, \cite{Shafi:2002qh} to  atmospheric and solar neutrino oscillations
(also called neutrino anarchy) assumes that all couplings involving the neutrino flavors are comparable in size.
This implies that large mixing angles
could be natural in this setting, which can be successfully realized with ${\cal U}(1)$
flavor symmetry.
The latter's role is also important for generating the hierarchies observed in charged fermion
masses and the CKM mixing elements
(generated by distinct transformations of  right-handed leptons and quarks).  The left-handed lepton doublets all
have the same transformation properties, i.e. $Q[l_1]=Q[l_2]=Q[l_3]$,
giving rise to large $\te_{12}$ and $\te_{23}$. However, by the same reasoning a priori
one expects to have large (not suppressed) $\te_{13}$. Also, the hierarchy
in (\ref{2nmasrat}) is not understood.
The latter, with the hierarchical mass spectrum, means that
$\fr{m_2}{m_3}\sim 0.2$. It is possible to get such a mass ratio within a democratic scenario if we arrange that
at leading level, the neutrino mass matrix
has only one massive eigenstate. With additional small perturbations, the mass of the second
 eigenstate will be somewhat suppressed. Within the see-saw scenario, this
can be fulfilled by the single RHN dominance \cite{Suematsu:1996mk}.
This mechanism  works well
within various GUTs augmented with ${\cal U}(1)$ flavor symmetry
\cite{Shafi:1999rm}, \cite{Shafi:1999au}.
As far as $\te_{13}$ is concerned, its suppression requires
more effort. It is possible to introduce additional discrete symmetries
\cite{Shafi:2004jy} which guarantee a suppressed value of $\te_{13}$.
However, it is difficult to realize such a possibility  in GUTs.
Thus, in the framework of
grand unification with ${\cal U}(1)$ flavor symmetry it is challenging to have bi-large neutrino mixings,
keep $\te_{13}$ sufficiently suppressed,  and also
explain the ratio $\fr{m_2}{m_3} \sim \lam$.
In this paper we will take up this challenge and develop a framework which is
also applicable to GUTs such as
$SO(10)$ \cite{Shafi:2005rd} and flipped $SU(5)$ \cite{Shafi:2006dm}. An
important role will be played by the double see-saw
mechanism which offers new ways for successful model building. For example,
the flipped $SU(5)$ GUT of \cite{Shafi:2006dm} gives $\te_{13} \approx 0.075$.

\section{Neutrino Textures and a Prediction of $\te_{13}$}

\subsection{Democracy with controlled zeros}

Consider a `light' neutrino texture
consisting of  two parts,
\beq
m_{\nu }=m_{\nu }^{(0)}+m_{\nu }^{(1)}~,
\la{mnu}
\eeq
where the leading part $m_{\nu }^{(0)}$ will be responsible for the
atmospheric neutrino anomaly, while the sub-leading entry $m_{\nu }^{(1)}$
will ensure large angle solar neutrino oscillations.
We will assume that
\beq
\begin{array}{ccc}
 & {\begin{array}{ccc}
\hs{-1cm} &  &\,\,~~~
\end{array}}\\ \vspace{2mm}
\!\!\!\!\! &~ {m_{\nu }^{(0)}
= \left(\hs{-0.2cm}\begin{array}{ccc}
~0&~0 &~0
\\
~0 & ~\al^2 &~ \al
\\
~0 &~ \al &~1
\end{array}\hs{-0.15cm}\right)\fr{m}{1+|\al |^2}~,~~~~~ }
{\rm with}~~~~\al \sim 1~,
\end{array}
\la{m0} \eeq
while
\begin{equation}
\begin{array}{cc}
& {\begin{array}{ccc}
\hs{-1.2cm}~ &~  &~
\end{array}}\\ \vspace{2mm}
\begin{array}{c}
  \\ \\

\end{array} &{m_{\nu }^{(1)}=\left(\begin{array}{ccc}
\bar \al^2 &~~~\bar \al &~~~0
\\
\bar \al  &~~~1&~~~0
\\
0  &~~~0 &~~~0
\end{array}\hs{-0.2cm}\right)m' ~,~~~~
}{\rm with}~~~~\bar \al \sim 1~.
\end{array}
\label{m1}
\end{equation}
The two matrices contain $2\tm 2$ sub-matrices with democratic structure
and zero determinants. We can thus refer to democratic  textures
 with controlled zeros\footnote{Later on,  these textures  will be justified by  symmetry arguments.}.
Next we assume that
\beq
m\gg m'~.
\la{solatmhier}
\eeq
This structure of $m_{\nu }$ already points to a hierarchical neutrino mass
spectrum. The leading eigenvalue of $m_{\nu }$
is close to the single non-zero eigenvalue of (\ref{m0}), namely
$m_{\nu_3}\simeq m$. Thus, the scale $m$ is relevant for  atmospheric
neutrino oscillations. Moreover, the 2-3 mixing angle $\te_{23}$ is
naturally large. The second non-zero eigenvalue of $m_{\nu }$ is determined
by the scale $m'$ responsible for solar neutrino oscillations.
The angle $\te_{12}$ is also naturally large.
The $1-2$ and $2-3$ rotations also induce a non-zero
mixing $\te_{13}$. A limited number of the parameters in the neutrino mass
matrix allows one to calculate the $\te_{13}$ in terms of other oscillation
parameters. In refs.  \cite{Frampton:2002qc, Ibarra:2003xp}, by invoking
two RHNs and specific neutrino
Dirac Yukawa texture (in a basis of diagonal RHN mass matrix)
a prediction for $\te_{13}$ was obtained. One of our main goals is to provide a derivation of the
textures in (\ref{m0}), (\ref{m1}) based on symmetry arguments.
Before doing this, let us  give an estimate for $\te_{13}$.

The leading part of $m_{\nu }$ can be diagonalized by the
transformation
$U_{23}^Tm_{\nu }^{(0)}U_{23}=\ov{m}_{\nu }^{(0)}={\rm Diag}\l 0,~0,~m\r $,
with
\beq
\begin{array}{ccc}
 & {\begin{array}{ccc}
 & &
\end{array}}\\ \vspace{2mm}
U_{23}=
\begin{array}{c}
\end{array}\!\!\!\!\!\! &P{\left(\begin{array}{ccc}
\hs{-0.1cm}1~,&
0 ~, &0
\\
\hs{-0.2cm}0~, &
\hs{-0.1cm}c_{23}~,& s_{23}
\\
\hs{-0.1cm} 0~,&-s_{23} ~,&
c_{23}
\end{array} \right)\! }~,
\end{array}
\la{U23}
\eeq
where $c_{23}\equiv \cos \te_{23}$, $s_{23}\equiv \sin \te_{23}$, and
\beq
\tan \te_{23}=|\al |~,~~~~
P={\rm Diag}\l 1,~e^{{\rm i} \chi },~1 \r ~,~~~
\chi =-{\rm Arg}(\al )~.
\label{PU23}
\end{equation}
By this rotation, the sub-leading part transforms into
\begin{equation}
\begin{array}{ccc}
 & {\begin{array}{ccc}
~ & &\,\,~~~
\end{array}}\\ \vspace{2mm}
\begin{array}{c}
 \\  \\
\end{array}\!\!\!\!\! &\ov{m}_{\nu }^{(1)}=U_{23}^Tm_{\nu }^{(1)}U_{23}
= \tl{P}\cdot {\left(\begin{array}{ccc}
\bar \al^2 & c_{23}\bar \al   & s_{23}\bar \al
\\
c_{23}\bar \al  & c_{23}^2 & c_{23}s_{23}
\\
s_{23}\bar \al & c_{23}s_{23} & s_{23}^2
\end{array}\right)\cdot \tl{P}m' } ~,
\end{array}
\hs{-1cm}
\label{mnu0}
\end{equation}
where
$\tl{P}={\rm Diag}(1, ~e^{-{\rm i} \chi },~e^{-{\rm i} \chi })$.

The matrix
$\ov{m}_{\nu }=\ov{m}_{\nu }^{(0)}+\ov{m}_{\nu }^{(1)}$
is diagonalized by the transformation
$U_{12}^TU_{13}^T\ov{m}_{\nu }U_{13}U_{12}=m_{\nu }^{\rm diag}$, where
$$
\begin{array}{ccc}
 & {\begin{array}{ccc}
 & &
\end{array}}\\ \vspace{2mm}
U_{12}\simeq P'\cdot
\begin{array}{c}
\end{array}\!\!\!\!\!\! &{\left(\begin{array}{ccc}
\hs{-0.1cm}c_{12}~,&
s_{12} ~, &0
\\
\hs{-0.2cm}-s_{12}~, &
\hs{-0.1cm}c_{12}~,& 0
\\
\hs{-0.1cm} 0~,&0 ~,&
1
\end{array} \right)\! }~,
\end{array}
~~~
\begin{array}{ccc}
 & {\begin{array}{ccc}
 & &
\end{array}}\\ \vspace{2mm}
U_{13}\simeq
\begin{array}{c}
\end{array}\!\!\!\!\!\! &{\left(\begin{array}{ccc}
\hs{-0.1cm}c_{13}e^{{\rm i}\cdot \de }~,&
0 ~, &s_{13}e^{{\rm i}\cdot \de }
\\
\hs{-0.2cm}0~, &
\hs{-0.1cm}1~,& 0
\\
\hs{-0.1cm}-s_{13}~,&0 ~,&
c_{13}
\end{array} \right)\! }~,
\end{array}
$$
\beq
P'={\rm Diag}\l e^{{\rm i}\cdot \chi'},~e^{-{\rm i}\cdot \chi },~
e^{-{\rm i}\cdot \chi } \r ~,~~~\chi'=-{\rm Arg}(\bar \al )~,
~~~\de =-{\rm Arg}\l \fr{m'\bar \al }{m}\r ~,
\label{U12}
\end{equation}
\beq
\tan \te_{12}=\fr{|\bar \al |}{c_{23}}~,~~~~~
\tan \te_{13}^{\nu }=\left |\fr{m'}{m}\right |s_{23}|\bar \al |~,
\la{tans}
\eeq
\beq
m_{\nu }^{\rm diag}={\rm Diag}(0, m_2, m)~,~~~~~ m_2=m'(|\bar \al |^2+c_{23}^2)~.
\la{mnudiag}
\eeq
 Using (\ref{tans}), (\ref{mnudiag}) and
the relation $\left |\fr{m_2}{m}\right |\simeq \sq{\fr{\De m_{\rm
sol}^2}{\De m_{\rm atm}^2}}$, we find \footnote{A similar suppression
$\te_{13}\sim \fr{m_{\rm sol}}{m_{\rm atm}}$ was obtained in a
democratic scenario \cite{Shafi:2004jy} with discrete symmetries.
We note, however, that it is hard to realize  this within a
GUT  supplemented by symmetry arguments.
For some recent works with flavor symmetries and suppressed $\theta_{13}$ see \cite{Rodejohann:2005ru}.
For a review of the models and
list of references, see \cite{Albrow:2005kw},
\cite{Alb}.}
 \beq
 \tan \te_{13}^{\nu
}\simeq \sq{\fr{\De m_{\rm sol}^2}{\De m_{\rm atm}^2}} \fr{\tan
\te_{12}\tan \te_{23}}{1+\tan^2 \te_{12}}~,
 \la{13pred}
 \eeq
where we have introduced the superscript
'$\nu $' in order to indicate that this contribution comes from the
neutrino sector\footnote{The physical leptonic mixing angle
$\te_{13}\equiv |U_{e3}^l|$ can receive  sizable contribution also from
the charged lepton sector.}.
Using the current data \cite{Fogli:2005gs}, we find
 \beq
\te_{13}^{\nu }\simeq 0.05-0.14~.
 \la{13range}
  \eeq
{} For the central values given in (\ref{data}), (\ref{13pred})
yields $(\te_{13}^{\nu })^{\rm cent}\simeq 0.075$.

It is possible to consider a slightly modified texture which yields a similar
(but slightly modified) relation. Namely, we leave the leading part $m_{\nu }^{(0)}$ unchanged, but for the
sub-leading part take
\begin{equation}
\begin{array}{cc}
& {\begin{array}{ccc}
\hs{-1.2cm}~ &~  &~
\end{array}}\\ \vspace{2mm}
\begin{array}{c}
  \\ \\

\end{array} &{m_{\nu }^{(1)}=\left(\begin{array}{ccc}
\bar \al^2 &~~~0 &~~~\bar \al
\\
0 &~~~0&~~~0
\\
 \bar \al  &~~~0 &~~~1
\end{array}\hs{-0.2cm}\right)m' ~,~~~~
}{\rm with}~~~~\bar \al \sim 1~.
\end{array}
\label{1m1}
\end{equation}
 We then find
\beq
 \tan \te_{13}^{\nu
}\simeq \sq{\fr{\De m_{\rm sol}^2}{\De m_{\rm atm}^2}} \fr{\tan
\te_{12}\cot \te_{23}}{1+\tan^2 \te_{12}}~,
 \la{13pred1}
 \eeq
which gives
\beq
\te_{13}^{\nu }\simeq 0.05-0.15~,~~~~~
(\te_{13}^{\nu })^{\rm cent}\simeq 0.094~.
 \la{13range1}
  \eeq

For these estimates we have assumed that the charged lepton sector practically
does not contribute to $\te_{12}$ and $\te_{23}$ mixings which are
determined largely from the neutral sector. This takes place in many GUT models
as we will discuss below.
In the (1,3) element of the lepton mixing matrix, if the charged
lepton sector also contributes, we can write
\beq
\te_{13}\equiv
|U_{e3}^l|\simeq \left | \tan \te_{13}^{\nu}-\ep_E
s_{23}e^{{\rm i}\rho }\right | \cos \te_{13}^{\nu }~,
\la{Ue3}
\eeq
where $\ep_E$ is related to the rotations coming from the charged lepton sector, and
$\rho $ is some (undetermined) phase.
We have parameterized the third mixing angle in this form (\ref{Ue3}) for convenience, which should become clear
later on.
With little or no information about the charged lepton sector we are  unable to
predict  $U_{e3}^l$. Thus,  it is important to take a close look at
 the charged fermion sector, and  GUTs are particularly interesting
from this viewpoint because they may closely relate charged and neutral sectors, thereby providing more information
on the size of $\ep_E$.
For instance, if  $\ep_E\ll 0.1$, then $\te_{13}$  is given by
$\te_{13}^{\nu }$ to a good approximation.

Before proceeding further with this, we first wish to have some understanding on how the textures (\ref{m0}) and (\ref{m1})
can be generated from a ${\cal U}(1)$ flavor symmetry.

\subsection{Deformed `Dirac' couplings and Double See-Saw}

We begin with the charged fermion sector
 aiming  to obtain  hierarchical masses and suppressed CKM mixings by
${\cal U}(1)$ symmetry. We introduce a MSSM singlet superfield $X$ with
${\cal U}(1)$ charge $Q[X]=-1$. The VEV of the scalar component of $X$ breaks
${\cal U}(1)$, and we assume that $\fr{\lan X\ran }{M_{\rm Pl}}\equiv
\lam  (\simeq 0.2)$.
With suitable ${\cal U}(1)$ charge prescriptions for the quark and lepton families,
we can realize desirable hierarchies expressed in powers of  $\lam$,
 an important expansion parameter which plays a central role in our construction.

The entries in the CKM matrix and the quark mass hierarchies
suggest the following ${\cal U}(1)$ charge assignments:
$$
Q[q_1]=a+3~,~~Q[q_2]=a+2~,~~Q[q_3]=a~,~~
$$
\beq
Q[u^c_1]=b+5~,~~Q[u^c_2]=b+2~,~~Q[u^c_3]=b~,~~
Q[d^c_1]=c+1~,~~Q[d^c_2]=Q[d^c_3]=c~,
\la{quarkCh}
\eeq
with $a, b, c$  not yet determined. For the
MSSM higgs doublets we have
\beq
Q[h_u]=-(a+b)~,~~~Q[h_d]=n-a-c~.
\la{hudCh}
\eeq
The Yukawa textures for up and down type quarks are respectively
\begin{equation}
\begin{array}{ccc}
 & {\begin{array}{ccc}
\hs{-1cm}u^c_1\hspace{0.8mm} & u^c_2
 &\hs{0.8mm} u^c_3
\end{array}}\\ \vspace{1mm}

\begin{array}{c}
q_1\vs{0.1cm} \\ q_2\vs{0.1cm} \\ q_3
 \end{array}\!\!\!\!\!\hs{-0.2cm} &{\left(\begin{array}{ccc}

 \lam^8 &\lam^5  & \lam^3
\\
 \lam^7 & \lam^4 & \lam^2
 \\
 \lam^5&\lam^4 & 1

\end{array}\right)h_u}~,~~~
\end{array}
\begin{array}{ccc}
 & {\begin{array}{ccc}
\hs{-1cm}d^c_1 &\hs{0.1cm} d^c_2 & \hs{0.1cm} d^c_3
\end{array}}\\ \vspace{1mm}

\begin{array}{c}
q_1 \\ q_2 \\ q_3
 \end{array}\!\!\!\!\!\hs{-0.2cm} &{\left(\begin{array}{ccc}

 \lam^4 &~\lam^3 &~ \lam^3
\\
 \lam^3 &~\lam^2 &~\lam^2
 \\
 \lam &~1&~ 1

\end{array}\right)\lam^nh_d}~.
\end{array}  \!\!
\label{Yud}
\end{equation}
It is easy to see that these textures indeed yield the  quark
Yukawa  hierarchies and CKM mixings given in (\ref{obsmass}).
An integer $n\geq 0$ determines the value of  $\tan \bt $.

Next we focus on the lepton sector.
Taking the assignments
$$
Q[l_1]=d+k~,~~Q[l_2]=d+m~,~~Q[l_3]=d~,~~
$$
\beq
Q[e^c_1]=a+c-d+r~,~~Q[e^c_2]=a+c-d+p~,~~Q[e^c_3]=a+c-d~,
\la{leptCh}
\eeq
the charged lepton Yukawa couplings are
\begin{equation}
\begin{array}{ccc}
 & {\begin{array}{ccc}
      \hs{-0.7cm}l_1 &\hs{0.5cm}
~l_2  &\hs{0.5cm}~ l_3
\end{array}}\\ \vspace{1mm}
\begin{array}{c}
e^c_1 \\ e^c_2 \\ e^c_3
 \end{array}\!\!\!\!\!\hs{-0.2cm} &{\left(\begin{array}{ccc}

 \lam^{k+r} & ~\lam^{m+r}  &~\lam^r
\\
 \lam^{k+p} &~\lam^{m+p} &~\lam^p
 \\
 \lam^k &~\lam^m  &~ 1

\end{array}\right)\lam^nh_d}~.
\end{array}  \!\!  ~~~~~
\label{Ye}
\end{equation}
The charged lepton mass hierarchies are reproduced if $m+p=2$ and $k+r=5$.

As far as the numbers $m, k, p, r$ are concerned, in the MSSM framework
they are not determined.
Thus, some comments are in order. If we take $m=k=0$ we have a democratic
scenario which would naturally yield large $\te_{12}$ and $\te_{23}$. However,
 problems due to large $\te_{13}$ and the ratio
$\sq{\De m_{\rm sol}^2/\De m_{\rm atm}^2}~$ arise,
and their resolution  especially within  GUTs is difficult.
Also, this choice would
reduce the success of the predictive texture presented in the previous
subsection.
Thus, we prefer a situation with
$m\geq 1$ and  $k>m$ (or a selection of $m$ and $k$ in such a way as to forbid
$(3.1)$ and $(3,2)$ entries in the matrix of (\ref{Ye})). With this choice
$\te_{12}$ and $\te_{23}$ are determined from the neutral sector
and a prediction for $\te_{13}$ with the  neutrino texture suggested earlier can work.
This option is also motivated from the viewpoint of GUTs because often in
models like $SO(10)$ and $E_6$, we expect  suppressed contributions from the
charged lepton sector. No other restrictions on these numbers are imposed for
the time being. (Note that if, for example, $k\geq 2+m$ the contribution to the
third mixing angle is suppressed and $\te_{13}$ is determined  from the
neutral lepton sector.)

Next let us introduce right-handed neutrinos. They are automatically present
in models which include a gauged $U(1)_{B-L}$ symmetry and/or left-right
symmetries such as $SO(10)$.
From $SU(2)_R$ we expect
\beq
Q[\nu^c_{\al }]=Q[e^c_{\al }]~,
\la{LRrel}
\eeq
and therefore neutrino Dirac masses will be induced from the couplings\footnote{For simplicity we will proceed with relation (\ref{LRrel}).
However, an unrestricted ${\cal U}(1)$ charge assignment for $\nu^c_i$ is possible.}

\begin{equation}
\begin{array}{ccc}
 & {\begin{array}{ccc}
\hs{-1.3cm}\nu^c_1 &
\hs{0.5cm}\nu^c_2  & \hs{0.5cm}\nu^c_3
\end{array}}\\ \vspace{1mm}
\begin{array}{c}
l_1 \\ l_2 \\ l_3
 \end{array}\!\!\!\!\!\hs{-0.2cm} &{\left(\begin{array}{ccc}

 \lam^{k+r} & ~\lam^{k+p}  &~\lam^k
\\
 \lam^{m+r} &~\lam^{m+p} &~\lam^m
 \\
 \lam^r &~ \lam^p &~ 1

\end{array}\right)\lam^{c-b}h_u}~,
\end{array}  \!\!  ~~~~~
\label{mD0}
\end{equation}
where we have assumed that $c-b\geq 0$ is integer.
With the Dirac mass matrix of (\ref{mD0}) and invoking the see-saw, one
can see that it is hard to get desirable bi-large neutrino oscillations
in a straightforward way.
Note also that in minimal $SO(10)$ or $E_6$ we have the relations
$m_D=M_U\propto M_D=M_E$.
Then, in (2,2) and (1,1) entries of $m_D, M_D, M_E$,
the exponents $\lam^4$ and $\lam^8$ respectively will remain.
These are unacceptably small contributions for the light down quarks and charged leptons.

To resolve these problems some extension is needed, with care
 taken
to avoid the relation $m_D=M_U$.
There are some
relations between neutral and charged sectors in GUTs, and the hope is to find
 hints on  how the neutrino Dirac couplings should be modified.
A simple extension of $SO(10)$, which cures problems in the charged
fermion sector, is to introduce  matter
$10$-plets \cite{Berezhiani:1996bv}, \cite{Shafi:1999au}
\cite{Maekawa:2001uk}.
For instance, in  the $SO(10)$ model of \cite{Shafi:2005rd}
this, together with improved quark and charged lepton sectors,
leads to deformed neutrino Dirac couplings.
The latter, as we will see shortly, is a crucial ingredient for generating the desirable texture.

We begin by introducing additional vector-like states which can modify
the neutrino Dirac mass matrix.
Introducing the vector-like lepton doublets $L, \bar L$  and
MSSM singlet scalar superfields $S, \bar S$ with ${\cal U}(1)$ charges
\beq
Q[L]=x~,~~Q[\bar L]=\bar x~,~~Q[S]=-(m+b-c+d+\bar x)~,~~
Q[\bar S]=-(p-b+c-d+x)~,
\la{LSch}
\eeq
the additional relevant couplings will be
\beq
\fr{\bar S}{M_{\rm Pl}}L\nu^c_2h_u+S\lam^{c-b}\bar Ll_2+
M_{\rm Pl}\lam^{x+\bar x}\bar LL~.
\la{LScoup}
\eeq
After substitution of appropriate VEVs we are dealing with the  mass
matrix
\begin{equation}
\begin{array}{cccc}
 & {\begin{array}{cccc}
\hs{-1cm}\nu^c_1 & \hs{0.8cm}
\nu^c_2 & \hs{0.8cm}\nu^c_3 & \hs{0.8cm}\bar L
\end{array}}\\ \vspace{1mm}

\begin{array}{c}
l_1 \\ l_2 \\ l_3  \\L
 \end{array}\!\!\!\!\!\hs{-0.2cm} &{\left(\begin{array}{ccccc}

 \lam^{k+r}h_u & \lam^{k+p}h_u  & \lam^kh_u  &0
\\
 \lam^{m+r}h_u &\lam^{m+p}h_u  &\lam^mh_u &\lan S\ran
 \\
 \lam^rh_u &\lam^ph_u & h_u &0
 \\
 0&\fr{\lan \bar S\ran h_u}{M_{\rm Pl}\lam^{c-b}} &0
&M_{\rm Pl}\fr{\lam^{x+\bar x}}{\lam^{c-b}}

\end{array}\right)\lam^{c-b}}~.
\end{array}  \!\!  ~~~~~
\label{ML}
\end{equation}
With
\beq
\fr{\lan S\bar S\ran }{M_{\rm Pl}^2\lam^{x+\bar x}}=\lam^{m_1}~,
\la{VEVfix}
\eeq
and assuming $m_1<m+p$,  integration of
 $L, \bar L$ states  reduces (\ref{ML}) to the following
$3\tm 3$ matrix
\begin{equation}
\begin{array}{ccc}
 & {\begin{array}{ccc}
\hs{-1.2cm}\nu^c_1 &
~\hs{0.4cm}\nu^c_2  &~\hs{0.4cm} \nu^c_3
\end{array}}\\ \vspace{1mm}
m_D\simeq
\begin{array}{c}
l_1 \\ l_2 \\ l_3
 \end{array}\!\!\!\!\!\hs{-0.2cm} &{\left(\begin{array}{ccc}

 \lam^{k+r} & ~\lam^{k+p}  &~\lam^k
\\
 \lam^{m+r} &~\lam^{m_1} &~\lam^m
 \\
\lam^{r}  &~\lam^p  &~ 1

\end{array}\right)\lam^{c-b}h_u}~.
\end{array}  \!\!  ~~~~~
\label{mD}
\end{equation}
Thus, the effective Dirac  mass matrix is modified, with $L, \bar L$ states playing an important role.
Note that there is  $l_2$-$L$ mixing in
(\ref{ML}),
\beq
L\supset \fr{\lan S\ran }{M_{\rm Pl}}\lam^{b-c-x-\bar x}l_2~.
\la{Ll2}
\eeq
This will be important for the sub-leading part of the neutrino mass matrix.

From the Dirac mass matrix (\ref{mD}), we  see that a priori the value
of $\te_{23}$ mixing angle is expected to be small ($\sim \lam^m$) because it is determined by the ratio of
$(2,3)$ and (3,3) elements. However, imagine that the state $\nu^c_3$ decouples by forming a very heavy state with
another MSSM singlet
$N_3$. Then $\te_{23}$ will be determined by the ratio of (2,2) and (3,2)
elements, and if we select $m_1=p$, $\te_{23}$ can be large. Thus, there is no need to decouple the  state $\nu^c_2$.
However, in order to have
 large $\te_{12}$ mixing and  zero (or appropriately suppressed) entries
in the neutrino texture, the state $\nu^c_1$ must decouple. It is possible
to realize all of this by introducing three additional GUT singlets, $N_i$
($i=1,2,3$). With  suitable couplings between these and the $\nu^c$ states, the desired goal can be achieved.
 This mechanism was discussed in \cite{Shafi:1998dv} and
applied within various GUTs \cite{Shafi:1999au}, \cite{Shafi:1998jf}.
Here we will use this idea for generating
the desired dominant part of neutrino  texture.

Assigning the following ${\cal U}(1)$ charges to $N_i$ states
\beq
Q[N_1]=-(a+c-d+r)~,~~~Q[N_2]=0~,~~~Q[N_3]=-(a+c-d)~,
\la{Nch}
\eeq
and, for simplicity, assuming that $r$ is non-integer [but $r+k$
is still integer, so the (1,1) element in (\ref{Ye}) is kept],
the relevant couplings are
\begin{equation}
\begin{array}{ccc}
 & {\begin{array}{ccc}
\hs{-1cm}N_1\hspace{0.8mm} & N_2
 &\hs{0.8mm} N_3
\end{array}}\\ \vspace{1mm}
M=
\begin{array}{c}
\nu^c_1\vs{0.1cm} \\ \nu^c_2\vs{0.1cm} \\ \nu^c_3
 \end{array}\!\!\!\!\!\hs{-0.2cm} &{\left(\begin{array}{ccc}

 1 &0  & 0
\\
 0&\lam^{a+c-d+p} &\lam^p
 \\
 0& \lam^{a+c-d} & 1

\end{array}\right)M_{\rm Pl}}~,~~~
\end{array}
\begin{array}{ccc}
 & {\begin{array}{ccc}
\hs{-0.7cm}N_1 &\hs{-0.1cm} N_2 & \hs{-0.1cm}N_3
\end{array}}\\ \vspace{1mm}
M_N=
\begin{array}{c}
N_1 \\ N_2 \\ N_3
 \end{array}\!\!\!\!\!\hs{-0.2cm} &{\left(\begin{array}{ccc}

 0 &~0 &~ 0
\\
 0&~1 &~0
 \\
 0&~0&~ 0

\end{array}\right)M_{\rm Pl}}~.
\end{array}  \!\!
\label{Ncoupl}
\end{equation}
The states $\nu^c_1$ and $\nu^c_3$ pair up with $N_1$ and
$N_3$ respectively and decouple. However, although $\nu^c_2$
couples with $N_2$ it does not decouple because $N_2$ itself has a large mass.

From (\ref{mD}) and (\ref{Ncoupl}), the $9\tm 9$ mass matrix for neutral
fermions is given by
\begin{equation}
\begin{array}{ccc}
 & {\begin{array}{ccc}
\hs{-0.15cm}\nu \hspace{0.2cm} & \hspace{0.15cm}
\nu^c \hspace{0.2cm}
& \hspace{0.15cm}N
\end{array}}\\ \vspace{1mm}

\begin{array}{c}
\nu \\ \nu^c \\ N
 \end{array}\!\!\!\!\!\hs{-0.2cm} &{\left(\begin{array}{ccccc}

 0 & m_D &0
\\
 m_D^T&0 &M
 \\
 0&M^T & M_N
\end{array}\right)}~.
\end{array}  \!\!  ~~~~~
\label{nubig}
\end{equation}
Integration of the heavy $\nu^c$, $N$ states leads to the light neutrino mass
matrix  given by the double see-saw formula:
\beq
m_{\nu }^{(0)}=m_D\fr{1}{M^T}M_N\fr{1}{M}m_D^T~.
\la{doubleSS}
\eeq
{} Taking $m_1=p$, one can  verify that
\beq
\begin{array}{ccc}
 & {\begin{array}{ccc}
\hs{-0.1cm} &
  &
\end{array}}\\ \vspace{1mm}
m_{\nu }^{(0)}\simeq
\begin{array}{c}
 \\  \\
 \end{array}\!\!\!\!\!\hs{-0.2cm} &{\left(\begin{array}{ccc}

 0 & 0  &0
\\
 0 &\al^2  &\al
 \\
 0 &\al   & 1

\end{array}\right)\fr{m}{1+|\al |^2}}~,
\end{array}  \!\!
\la{derm0}
\eeq
where $\al\sim 1$ is a dimensionless coupling and
$m\sim \fr{\lan h_u^{(0)}\ran^2}{M_{\rm Pl}\lam^{2(a+b-d)}}$.
This is indeed the form which we required for the leading
contribution to $m_\nu$.

As far as the sub-leading part is concerned, it can be
generated through an additional singlet state ${\cal N}$. Since the additional part
will be generated by a single state, its degeneracy will be due to the `dominance' mechanism in \cite{Suematsu:1996mk}.
In our case, an important role will be played by the state $L$ which mixes with $l_2$.
With a ${\cal U}(1)$ charge prescription $Q[{\cal N}]=a+c-d-k+y$,
the relevant couplings are
\beq
\lam^{c-b+y}l_1{\cal N}h_u+\lam^{2d-p-k+y}\fr{\bar S}{M_{\rm Pl}}L{\cal N}h_u+
M_{\rm Pl}\lam^{2(a+c-d-k+y)}{\cal N}^2~.
\la{calN}
\eeq
Using (\ref{VEVfix}), (\ref{Ll2}) and taking $d=c+k/2$, from (\ref{calN})
the effective couplings are
\beq
\lam^{c-b+y}l_1{\cal N}h_u+\lam^{c-b+y}l_2{\cal N}h_u+
M_{\rm Pl}\lam^{2(a+c-d-k+y)}{\cal N}^2~.
\la{calN1}
\eeq
Integrating out ${\cal N}$, from the couplings
in (\ref{calN1}) we can generate $m_{\nu }^{(1)}$ which has the form given
in (\ref{m1}), with
$m'\sim \fr{\lan h_u^{(0)}\ran^2\lam^{2k} }{M_{\rm Pl}\lam^{2(a+b-d)}}$.
With $k=1/2$ we obtain
\beq
\fr{m'}{m}\sim \lam ~.
\la{solRatAtm}
\eeq
Note also that with this selection of $k$, the $(3,1)$, $(3,2)$ elements
in (\ref{Ye}) vanish. Thus, $\te_{13}$ acquires contributions only from
the neutrino sector.

We have therefore generated a  neutrino texture which yields bi-large neutrino
mixing and predicts the value of $\te_{13}$. In addition, from
${\cal U}(1)$ charge assignments, we obtain $\fr{m'}{m}\sim \lam $, which
nicely accounts for the hierarchy in (\ref{2nmasrat}).

\section{GUTs and the Value of $\te_{13}$ }

\subsection{$SO(10)$ Example}

The mechanism presented in the previous section can be successfully realized
within various GUTs. In particular, the $SO(10)$ model of \cite{Shafi:2005rd} has all the ingredients
needed for the generation of neutrino textures of (\ref{mnu})-(\ref{m1}).
In addition, the 1-2 mixing coming from the charged
lepton sector is predicted to be $\te_{e\mu }\simeq \sq{\fr{m_e}{m_{\mu }}}$.
This gives a sizable
contribution to
$\te_{13}\equiv |U_{e3}^l|$, for which we obtain
\beq
\te_{13}\equiv
|U_{e3}^l|\simeq \left | \tan \te_{13}^{\nu}- \sq{\fr{m_e}{m_{\mu
}}}s_{23}e^{{\rm i}(\de -\chi )}\right | \cos \te_{13}^{\nu }~,
\la{so10Ue3}
\eeq
where $\te_{13}^{\nu}$ angle is defined in (\ref{13pred}), while
the phase $\de -\chi $ is unknown.
We can estimate the possible range for $|U_{e3}^l|$. Namely, from (\ref{so10Ue3}),
$$
{\rm for}~~~\te_{13}^{\nu }=0.05~,~~~~\te_{13}=0.01-0.09~,
$$
\beq
{\rm for}~~~\te_{13}^{\nu }=0.14~,~~~~\te_{13}=0.08-0.2~.
\la{e3ranges}
\eeq
Thus,  we find
\beq
\te_{13}=0.01-0.2~.
\la{Ue3SO10GUT}
\eeq
The upper range in  (\ref{Ue3SO10GUT}) is consistent with the
current experimental bound \cite{Apollonio:2002gd}, while future
experiments \cite{Itow:2001ee}, \cite{Albrow:2005kw} should be able to probe
values close to $0.01$.

\subsection{Flipped $SU(5)$ GUT}

Within flipped $SU(5)$ GUT it is also possible to generate the
desirable neutrino texture  (see  \cite{Shafi:2006dm} for details).
Among  other salient
features, flipped $SU(5)$ predicts the following mixings
from the charged lepton sector,   $\te_{e\mu }\sim \lam^3$
and $\te_{e\tau }\sim \lam^5$.
These  contributions  can be safely ignored
in estimating
$\te_{13}(\equiv |U_{e3}^l|)$ which is then completely determined from the
neutrino sector. According to (\ref{13pred}),
\beq
\te_{13} \simeq \te_{13}^{\nu}=0.05-0.14~~~~
(\mbox{with central value close to 0.075}).
\la{Ue3flipGUT}
\eeq

\vspace{0.8cm}
Below we discuss the leptonic rare decays and the Leptogenesis. These issues, within scenarios
leading to similar neutrino mass matrix, have been discussed in \cite{Guo:2006rm} and
\cite{Frampton:2002qc, Ibarra:2003xp, Guo:2003cc} respectively. However,
since these constructions are based on ad-hoc coupling textures, the angle $\theta_{13}$ is a free
parameter (no contribution from the charged lepton sector was discussed). Our consideration, with well
defined framework with ${\cal U}(1)$ flavor symmetry, predicting $\theta_{13}$ enables one to reduce number
of parameters and increase the predictive power.

\section{Rare Decays: $l_i\to l_j\ga $}

The see-saw mechanism which we have used for neutrino mass generation also
provides a source for lepton flavor violating
rare processes such as
$l_{\al } \to l_{\bt }\ga $ \cite{Borzumati:1986qx}.
We assume flavor universality at M$_{GUT}$ scale. Thus, flavor violation
 may arise if one or more of the RHN's mass is below  M$_{GUT}$.
In the scenario we presented in the previous sections such candidates are
$\nu^c_2$ and ${\cal N}$.
After integrating out the other heavy RHNs [from (\ref{Ncoupl}), (\ref{nubig}), (\ref{calN1})],
we  have the following effective couplings
\beq
h_ul^T ~\lam_D  \hs{-0.3cm}
\begin{array}{cc}
\vspace{1mm}

\begin{array}{c}
\vs{0.1cm}\\
 \end{array}\!\!\!\!\!\hs{-0.02cm} &{\left(\hs{-0.15cm}
 \begin{array}{cc}
{\cal N}
\\
\nu^c_2

\end{array}\hs{-0.15cm}\right)}\vs{-0.2cm}~
\end{array}
+~({\cal N}, \nu^c_2)~\hat{M}_R\hs{-0.3cm}
\begin{array}{cc}
\vspace{1mm}

\begin{array}{c}
\vs{0.1cm}\\
 \end{array}\!\!\!\!\!\hs{-0.02cm} &{\left(\hs{-0.15cm}
 \begin{array}{cc}
{\cal N}
\\
\nu^c_2

\end{array}\hs{-0.15cm}\right)}\vs{-0.2cm}~,
\end{array}
\label{lept-D-M}
\end{equation}
with
\beq
\begin{array}{ccc}
 & {\begin{array}{cc}
\hs{-0.2cm} \hspace{0.8cm} & \hspace{0.5cm}
\end{array}}\\ \vspace{1mm}
\lam_D=
\begin{array}{c}
 \vs{0.1cm}\\ \vs{0.1cm} \\
 \end{array}\!\!\!\!\!\hs{-0.2cm} &{\left(\begin{array}{ccc}

\hs{0.2mm} \bar{\lam }_1&\hs{0.2cm} 0
\\
\vs{-0.3cm}
\\
 \hs{0.2mm}\bar{\lam }_2& \hs{0.2cm}\lam_1
 \\
 \vs{-0.3cm}
 \\
 \hs{0.2mm}0 \hs{0.1cm}&\hs{0.2cm}\lam_2

\end{array}\right)},~~~~
\end{array}
{\hat M}_R={\rm Diag}\l M_1~,~ M_2\r ~,
\label{Dir-Maj}
\end{equation}
where
$$
M_1=M_{\cal N}\sim M_{\rm Pl}\lam^{2(a+c-d+y)-1}~,~~~~M_2=M_{\nu^c_2}\sim M_{\rm Pl}\lam^{2(a+c-d+p)}~,
$$
\beq
\lam_1\sim \lam_2\sim \lam^{c-b+p}~,~~~~\bar{\lam }_1\sim \bar{\lam }_2\sim \lam^{c-b+y}~.
\la{proport}
\eeq
One can easily
check out that the neutrino mass matrix $m_{\nu }=(h_u^0)^2\lam_D\fr{1}{\hat{M}_R}\lam_D^T$ derived from
(\ref{lept-D-M}), (\ref{Dir-Maj}) coincides with the texture given in
(\ref{mnu})-(\ref{m1}), with the following relations
\beq
\lam_1/\lam_2=\al ~,~~~\bar{\lam }_1/\bar{\lam }_2=\bar{\al }~,~~~\fr{\lam_2^2}{M_2}(h_u^0)^2=\fr{m}{1+|\al |^2}~,~~~
\fr{\bar{\lam }_2^2}{M_1}(h_u^0)^2=m' ~.
\la{Rarerels}
\eeq
The (1,2) element of $\lam_D$ vanishes because we have taken $k=1/2$.

 Assuming $N=1$ supergravity and universality of soft scalar
masses at $M_{\rm GUT}$, the non-universal contributions to the slepton masses are
\beq
(\de m^2)_{ij}\approx -\fr{1}{8\pi^2}(A+3)m_S^2T_{ij}~,~~{\rm with}~~
T_{ij}=
\sum_{\al =1,\hs{0.1cm}2}(\lam_D)_{i\al}
\ln \fr{M_{\rm GUT}}{M_{\al }}(\lam_D^T)_{j \al }~.
\la{insert1}
\eeq
From (\ref{insert1}), (\ref{Dir-Maj}) one finds
\beq
T_{12}= \bar{\lam }_1\bar{\lam }_2\ln \fr{M_{\rm GUT}}{M_1}~,~~~~
T_{23}=  \lam_1\lam_2 \ln \fr{M_{\rm GUT}}{M_2}~,~~~~
T_{13}= 0~.
\la{insert2}
\eeq
Therefore one expects no $\tau \to e\ga $ transition. In the models of
\cite{Shafi:2005rd}, \cite{Shafi:2006dm} it does not vanish but is strongly
suppressed.
For the other two channels, taking into account
eqs. (\ref{Rarerels}),  (\ref{insert2}) and the expressions given in \cite{Hisano:1995cp},
we have
$$
{\rm BR}(\mu \to e\ga )=
\fr{\al_{em}^3[(\de m^2)_{12}]^2}{G_F^2m_S^8}\tan^2 \bt =
\fr{\al_{em}^3(A+3)^2}{64\pi^4G_F^2m_S^4}
\l \bar{\lam }_1\bar{\lam }_2\r^2\tan^2 \bt \ln^2 \fr{M_{\rm GUT}}{M_1}=
$$
\beq
\fr{\al_{em}^3(A+3)^2}{64\pi^4G_F^2m_S^4}
\l \fr{1+\tan^2\bt }{\tan \bt }\r^2
\l \fr{s_{12}c_{12}}{c_{23}}\fr{M_1\sq{\De m_{\rm sol}^2}}{v^2}\r^2\ln^2 \fr{M_{\rm GUT}}{M_1}~,
\la{BRm-ega}
\eeq
and
$$
\fr{\rm BR(\tau \to \mu \ga )}{{\rm BR}(\mu \to e\ga )}\simeq
\l \fr{s_{23}}{s_{12}c_{12}(1+\tan^2\te_{12}+\tan^2\te_{23})}\r^2 \fr{\De m_{\rm atm}^2}{\De m_{\rm sol}^2}
\l \fr{M_2}{M_1}\r^2\l \fr{\ln M_{\rm GUT}/M_2}{\ln M_{\rm GUT}/M_1}\r^2\simeq
$$
\beq
12 \cdot \l \fr{M_2}{M_1}\r^2\l \fr{\ln M_{\rm GUT}/M_2}{\ln M_{\rm GUT}/M_1}\r^2~.
\la{Brs}
\eeq
{}For $A\sim 1$ we find
\beq
{\rm BR}(\mu \to e\ga )\simeq 1.2\cdot 10^{-11}\l \fr{250 {\rm GeV}}{m_S}\r^4
\l \fr{M_1}{2\cdot 10^{11}~{\rm GeV}}\r^2
\l \fr{\ln M_{\rm GUT}/M_1}{12.2}\r^2 ~,~~{\rm for}~~\tan \bt \simeq 60~,
\la{Brmue}
\eeq
where we have  selected the parameters such that the current experimental
bound ${\rm BR}(\mu \to e\ga)\stackrel{<}{_\sim }1.2\cdot 10^{-11}$
\cite{Brooks:1999pu} is satisfied.
For $M_1\stackrel{<}{_\sim }4.5\cdot 10^9$~GeV the branching ratio (${\rm BR}(\mu \lesssim e\ga)<10^{-14}$) will be below
the sensitivity expected in future experiments.

For  $M_1\stackrel{>}{_\sim }4.5\cdot 10^9$~GeV, the process $\mu \to e\ga $ may be testable
and we have three possible characteristic signatures for rare decays.

{\bf 1)} $M_2>M_1$, which yields

\beq
{\rm BR}(\mu \to e\ga )\ll {\rm BR}(\tau \to \mu \ga )~.
\la{sign2}
\eeq
The branching ratio for $\tau \to \mu \ga $ can easily reach
the current experimental upper bound
${\rm BR}(\tau \to \mu \ga )\stackrel{<}{_\sim }6.8\cdot 10^{-8}$
\cite{Aubert:2005ye} for $M_2\sim 30 M_1$
(for parameters close to those given in eq. (\ref{Brmue})).
Thus, both  $\mu \to e\ga $ and $\tau \to \mu \ga $
are potentially testable. This region of parameter space is also favored by
leptogenesis which we discuss in the next section.

{\bf 2)} $M_2\sim M_1/4$, which gives

\beq
{\rm BR}(\mu \to e\ga )\sim {\rm BR}(\tau \to \mu \ga )~,
\la{sign1}
\eeq

{\bf 3)} $M_2<M_1/10$, such that

\beq
{\rm BR}(\mu \to e\ga )\gg {\rm BR}(\tau \to \mu \ga )~,
\la{sign3}
\eeq
which means that if the branching of $\mu \to e\ga $ approaches the experimental
bound ($\stackrel{<}{_\sim }1.2\cdot 10^{-11}$), then $\tau \to \mu \ga $
is expected to be well  below its present and future sensitivity($\sim 10^{-9}$).

Note that specific GUT scenarios  can be more restrictive as far as
 dominant rare decays are concerned. For example, the flipped $SU(5)$ GUT presented
in \cite{Shafi:2006dm} with  neutrino texture similar to those presented in the previous
sections suggests possibility {\bf 1)}.
Hopefully, the ongoing (and planned) experiments will soon
discriminate among  the various possibilities.

\section{Leptogenesis and Prediction for Neutrino CP Violation}

The generation of predictive neutrino texture in our scenario occurs
through the see-saw mechanism, and we now exploit this ingredient for leptogenesis \cite{Fukugita:1986hr}
We will see that our scheme allows us to relate  CP violation in
neutrino oscillations with the CP violation needed for leptogenesis.

For leptogenesis the two relevant right-handed neutrino states are ${\cal N}$ and $\nu^c_2$. The lightest RHN is responsible
for out of equilibrium decays, while $\nu_2^c$ appears through
the loop contribution to CP violation.
The effective couplings of these RHNs are given in (\ref{lept-D-M}), (\ref{Dir-Maj}).
It is possible to redefine the phases of appropriate states in such a way as to have real Majorana masses of two RHNs, and
bring $\lam_D$ to the form
\beq
\begin{array}{ccc}
 & {\begin{array}{cc}
\hs{-0.2cm} \hspace{0.8cm} & \hspace{0.5cm}
\end{array}}\\ \vspace{1mm}
\lam_D\to
\begin{array}{c}
 \vs{0.1cm}\\ \vs{0.1cm} \\
 \end{array}\!\!\!\!\!\hs{-0.2cm} &{\left(\begin{array}{ccc}

\hs{0.2mm} |\bar{\lam}_1|&\hs{0.2cm} 0
\\
\vs{-0.3cm}
\\
 \hs{0.2mm}|\bar{\lam }_2|e^{{\rm i}\bar b}& \hs{0.2cm}|\lam_1 |
 \\
 \vs{-0.3cm}
 \\
 \hs{0.2mm}0 \hs{0.1cm}&\hs{0.2cm}|\lam_2|

\end{array}\right)}~.
\end{array}
\label{Dir-Maj1}
\end{equation}
As we see, thanks to the specific texture realized by ${\cal U}(1)$ flavor symmetry, in this system only
one (unremovable) phase appears\footnote{Leptogenesis by postulating a texture similar to (\ref{Dir-Maj}) was considered
in \cite{Frampton:2002qc}.}. The same (single) phase appears in the lepton mixing matrix.

With negligible contributions from the charged sector, the leptonic CP violation mainly arises from the neutrino
sector. The corresponding Jarlskog invariant
\beq
{\cal J}^l={\rm Im}\l U_{e2}^*U_{\mu 3}^*U_{e3}U_{\mu 2}\r
\la{jarls}
\eeq
measures the CP violation. Namely, we have
\beq
\sin 2\bar b=-\fr{4{\cal J}^l}{\te_{13}\sin 2\te_{12}\sin 2\te_{23}}~.
\la{sin-CP}
\eeq

The lepton asymmetry arises from the out of equilibrium decays of ${\cal N}$ and is given by
$\fr{n_L}{s}\simeq 10^{-3}\kappa_f\tl{\ep }_1$ \cite{Covi:1996wh},
where  the efficiency factor is $\kappa_f\simeq 2.4\cdot 10^{-2}$ in our model.
The  lepton asymmetry is partially converted to the observed baryon asymmetry via sphalerons \cite{Kuzmin:1985mm}. Namely,
$\fr{n_B}{s}=-\fr{8}{23}\fr{n_L}{s}$, so that
\beq
\fr{n_B}{s}\simeq -1.76\cdot 10^{-6}\tl{\ep }_1~.
\la{B-asym}
\eeq
The $\tl{\ep }_1$ comes from the interference of the tree and
one loop decays,
\beq
\tl{\ep }_1=-\fr{1}{2\pi }\l \sq{\xi }\ln \fr{1+\xi }{\xi }+\fr{2\sq{\xi }}{\xi -1}\r I_{21}~,~~~\xi =\fr{M_2^2}{M_1^2}~,
\la{tl-ep1}
\eeq

\beq
I_{21}=\fr{{\rm Im}[(\lam_D^{\dagger }\lam_D )^2_{21}]}{(\lam_D^{\dagger }\lam_D )_{11}}=\fr{|\lam_1|^2\sin 2\bar b}{1+\tan^2 \te_{12}c^2_{23}}~.
\la{I21}
\eeq
{}From the neutrino sector we have $|\lam_1 |^2/M_2\simeq \sq{\De m^2_{\rm atm}}s^2_{23}/(v\sin \bt )^2$,
and with  $\xi \gg 1$, we find from (\ref{tl-ep1})
\beq
\tl{\ep }_1=-\fr{3}{2\pi }\fr{M_1\sq{\De m^2_{\rm atm}}}{v^2}\l 1+\fr{1}{\tan^2 \bt }\r
\fr{s^2_{23}}{1+\tan^2\te_{12}c^2_{23}}\sin 2\bar b ~.
\la{tl-ep1-1}
\eeq
Combining (\ref{sin-CP}), (\ref{B-asym}), (\ref{tl-ep1-1}), using parameters from the neutrino data
and $\te_{13}\simeq 0.05-0.14$ (obtained from our texture), we finally  have
\beq
\fr{n_B}{s}\simeq -(1.7-4.2)\cdot 10^{-9}\l \fr{M_1}{10^{11}{\rm GeV}}\r \l 1+\fr{1}{\tan^2 \bt }\r {\cal J}^l~,
\la{B-asym-model}
\eeq
{}From the observed baryon asymmetry $\l \fr{n_B}{s}\r^{\rm exp}\simeq 9\cdot 10^{-11}$,
from (\ref{B-asym-model}) the leptonic CP violation is given by
\beq
{\cal J}^l=
\left\{ \begin{array}{lll}
-0.016; &~ {\rm for}~\tan \bt \sim 1 \\
-0.031; &~ {\rm for}~\tan \bt \stackrel{>}{_\sim }10
\end{array}
\right.~.
\la{J-pred}
\eeq
Future experiments should be able to probe CP violation of this magnitude.

Finally, note that for low values of $M_1=M_{\cal N}$, in order to obtain the required scale for
solar neutrino oscillation, the appropriate  Yukawa couplings must be suppressed. This will also
 suppress rare decay processes.
Thus, there is correlation between  CP violation [see (\ref{B-asym-model})]
and ${\rm BR}(\mu \to e\ga )$. In Table 1 we display these two quantities for different choices of
$M_1$. It is intriguing that for $M_1\approx 10^{11}-10^{12}$~GeV, both ${\cal J}^l$ and rare decays
 can be probed by experiments, and, as was pointed out in the previous section,
the decay $\tau \to \mu \ga $ dominates over $\mu \to e\ga $.

%
\begin{table} \caption{Values of ${\rm BR}(\mu \to e\ga )$ and ${\cal J}^l$ with $m_S\simeq 250$~GeV, $\tan \bt =60$
and  $n_B/s\simeq 9\cdot 10^{-11}$.}

\label{ginfl} $$\begin{array}{|c|c|c|}

\hline
\vspace{-0.3cm}
&& \\

M_1/{\rm GeV} & {\rm BR}(\mu \to e\ga )\tm 10^{11} &  {\cal J}^l\tm 10^2 \\
\vspace{-0.3cm}
&& \\

\hline \hline
\vspace{-0.4cm}

&& \\
 10^{11}&0.3  &-3.1 \\
\vspace{-0.4cm}
&& \\
\hline
\vspace{-0.3cm}
&& \\
 2\cdot 10^{11}& 1.2  &-1.6\\
\hline
\vspace{-0.3cm}
&& \\
 10^{12}& 22  &-0.31\\
\hline

\end{array}$$

\end{table}
%
%
%

\section{Conclusions}

 Neutrino oscillation
experiments have already proved the existence of new physics beyond the SM/MSSM, and the  planned long baseline
neutrino experiments \cite{Itow:2001ee}, \cite{Albrow:2005kw} should be able
to probe the value of
the third leptonic mixing angle $\te_{13}$  down to $\sim 10^{-2}$.
We have proposed  neutrino mass textures using double see-saw which allow one to
 predict the magnitude of $\te_{13}$. Our construction includes a
${\cal U}(1)$ flavor symmetry for understanding the hierarchies between fermion
masses and mixings.
In this setting we have also investigated
the rare decays, $l_i\to l_j\ga $, which arise from the right handed neutrino states below
the GUT scale. The process $\tau \to e\ga $ turns out to be either
absent or  strongly suppressed, while for  $\mu \to e\ga $ and $\tau \to \mu \ga $, we find the interesting possibility that
${\rm BR}(\mu \to e\ga )\sim {\rm BR}(\tau \to \mu \ga )$.

In implementing leptogenesis we uncover the intriguing possibility that the
lepton asymmetry is directly related to CP violation in neutrino oscillations.
Thus, the construction enables us to estimate two unknown quantities from the
neutrino sector, namely $\te_{13}$  and the magnitude of CP violation.

Within a  SUSY $SO(10)$  GUT we find
$\te_{13}\approx 0.01-0.2$, while in flipped $SU(5)$,
$\te_{13}\simeq 0.075$.

\vs{0.5cm}

\hs{-0.7cm}{\bf Acknowledgments}

\vs{0.2cm}
\hs{-0.7cm}Z.T. would like to thank the Bartol Research Institute for
warm hospitality during his visit.


\bibliographystyle{unsrt}

\begin{thebibliography}{99}




\bibitem{Fukuda:2000np}
  S.~Fukuda {\it et al.}  [Super-Kamiokande Collaboration],
  Phys.\ Rev.\ Lett.\  {\bf 85}, 3999 (2000)
  [hep-ex/0009001].
%
%
\bibitem{Fukuda:2001nj}
  S.~Fukuda {\it et al.}  [Super-Kamiokande Collaboration],
  Phys.\ Rev.\ Lett.\  {\bf 86} (2001) 5651
  [hep-ex/0103032];\\
K.~Eguchi {\it et al.}  [KamLAND Collaboration],
  Phys.\ Rev.\ Lett.\  {\bf 90} (2003) 021802
  [hep-ex/0212021].
%
\bibitem{Maltoni:2004ei}
  M.~Maltoni, T.~Schwetz, M.~A.~Tortola and J.~W.~F.~Valle,
  New J.\ Phys.\  {\bf 6} (2004) 122
  [hep-ph/0405172].


\bibitem{Fogli:2005gs}
  G.~L.~Fogli, E.~Lisi, A.~Marrone, A.~Palazzo and A.~M.~Rotunno,
  [hep-ph/0506307].



\bibitem{seesaw}
P.~Minkowski,
  Phys.\ Lett.\ B {\bf 67} (1977) 421;
  M.~Gell-Mann, P.~Ramond and R.~Slansky, in {it Supergravity} eds.
  P. van Nieuwenhuizen and D.Z. Freedman (North Holland, Amsterdam, 1979) p. 315;
  T.~Yanagida,
{\it In Proceedings of the Workshop on the Baryon Number of the Universe and Unified Theories, Tsukuba, Japan, 13-14 Feb 1979};
R.~N.~Mohapatra and G.~Senjanovic,
  Phys.\ Rev.\ Lett.\  {\bf 44} (1980) 912.





\bibitem{Apollonio:2002gd}
  M.~Apollonio {\it et al.},
  Eur.\ Phys.\ J.\ C {\bf 27} (2003) 331
  [hep-ex/0301017].




\bibitem{Froggatt:1978nt}
  C.~D.~Froggatt and H.~B.~Nielsen,
  Nucl.\ Phys.\ B {\bf 147} (1979) 277.


\bibitem{WMAP}
D.N. Spergel, et.al.,
{\it Three-Year Wilkinson Microwave Anisotropy Probe (WMAP) Observations:
Implications for Cosmology},
http://lambda.gsfc.nasa.gov/.


\bibitem{Fukugita:2006rm}
  M.~Fukugita, K.~Ichikawa, M.~Kawasaki and O.~Lahav,
  Phys.\ Rev.\  D {\bf 74} (2006) 027302
  [astro-ph/0605362].


\bibitem{Ellis:1999my}
  J.~R.~Ellis and S.~Lola,
  Phys.\ Lett.\ B {\bf 458} (1999) 310
  [hep-ph/9904279].


\bibitem{Wetterich:1998vh}
 C.~D.~Carone and M.~Sher,
  Phys.\ Lett.\ B {\bf 420} (1998) 83
  [hep-ph/9711259];\\
 C.~Wetterich,
  Phys.\ Lett.\ B {\bf 451} (1999) 397
  [hep-ph/9812426];\\
R.~Barbieri, L.~J.~Hall, G.~L.~Kane and G.~G.~Ross,
  hep-ph/9901228;\\
S.~Antusch and S.~F.~King,
  Nucl.\ Phys.\ B {\bf 705} (2005) 239
  [hep-ph/0402121];\\
S.~F.~King,
  JHEP {\bf 0508} (2005) 105
  [hep-ph/0506297]; See also references therein.




\bibitem{Petcov:1982ya}
  S.~T.~Petcov,
  Phys.\ Lett.\ B {\bf 110} (1982) 245;\\
R.~Barbieri, L.~J.~Hall, D.~R.~Smith, A.~Strumia and N.~Weiner,
  JHEP {\bf 9812} (1998) 017
  [hep-ph/9807235];\\
A.~S.~Joshipura and S.~D.~Rindani,
  Eur.\ Phys.\ J.\ C {\bf 14} (2000) 85
  [hep-ph/9811252];\\
R.~N.~Mohapatra, A.~Perez-Lorenzana and C.~A.~de Sousa Pires,
  Phys.\ Lett.\ B {\bf 474} (2000) 355
  [hep-ph/9911395];\\
W.~Grimus and L.~Lavoura,
  Phys.\ Rev.\ D {\bf 62} (2000) 093012
  [hep-ph/0007011].



\bibitem{Shafi:2000su}
  Q.~Shafi and Z.~Tavartkiladze,
  Phys.\ Lett.\ B {\bf 482} (2000) 145
  [hep-ph/0002150];\\
S.~F.~King and N.~N.~Singh,
  Nucl.\ Phys.\ B {\bf 596} (2001) 81
  [hep-ph/0007243].

\bibitem{Shafi:2001se}
  Q.~Shafi and Z.~Tavartkiladze,
  hep-ph/0101350.








\bibitem{Leontaris:2004rd}
  G.~K.~Leontaris, J.~Rizos and A.~Psallidas,
  Phys.\ Lett.\ B {\bf 597} (2004) 182
  [hep-ph/0404129];\\
  W.~Grimus and L.~Lavoura,
  J.\ Phys.\ G {\bf 31} (2005) 683
  [hep-ph/0410279];\\
  G.~Altarelli and R.~Franceschini,
  hep-ph/0512202.


\bibitem{Tanimoto:1999pj}
  M.~Tanimoto, T.~Watari and T.~Yanagida,
  Phys.\ Lett.\ B {\bf 461} (1999) 345
  [hep-ph/9904338];\\
L.~J.~Hall, H.~Murayama and N.~Weiner,
  Phys.\ Rev.\ Lett.\  {\bf 84} (2000) 2572
  [hep-ph/9911341];\\
N.~Haba and H.~Murayama,
  Phys.\ Rev.\ D {\bf 63} (2001) 053010
  [hep-ph/0009174].


\bibitem{Shafi:2002qh}
  Q.~Shafi and Z.~Tavartkiladze,
  Phys.\ Lett.\ B {\bf 550} (2002) 172
  [hep-ph/0208162]; Phys.\ Rev.\ D {\bf 67} (2003) 075007
  [hep-ph/0210181]; Nucl.\ Phys.\ B {\bf 665} (2003) 469
  [hep-ph/0303150];\\
G.~Altarelli, F.~Feruglio and I.~Masina,
  JHEP {\bf 0301} (2003) 035
  [hep-ph/0210342].



\bibitem{Antusch:2003kp}
  S.~Antusch, J.~Kersten, M.~Lindner and M.~Ratz,
  Nucl.\ Phys.\ B {\bf 674} (2003) 401
  [arXiv:hep-ph/0305273].



\bibitem{Suematsu:1996mk}
  D.~Suematsu,
  Phys.\ Lett.\ B {\bf 392} (1997) 413
  [hep-ph/9604242];\\
S.~F.~King,
  Phys.\ Lett.\ B {\bf 439} (1998) 350
  [hep-ph/9806440];\\
S.~Davidson and S.~F.~King,
  Phys.\ Lett.\ B {\bf 445}, 191 (1998)
  [hep-ph/9808296];\\
Q.~Shafi and Z.~Tavartkiladze,
  Phys.\ Lett.\ B {\bf 451}, 129 (1999)
  [hep-ph/9901243].






\bibitem{Shafi:1999rm}
  Q.~Shafi and Z.~Tavartkiladze,
  Phys.\ Lett.\ B {\bf 451} (1999) 129
  [hep-ph/9901243];
Nucl.\ Phys.\ B {\bf 573} (2000) 40
  [hep-ph/9905202].



\bibitem{Shafi:1999au}
  Q.~Shafi and Z.~Tavartkiladze,
  Phys.\ Lett.\ B {\bf 487} (2000) 145
  [hep-ph/9910314];
  Nucl.\ Phys.\ B {\bf 665} (2003) 469
  [hep-ph/0303150].



\bibitem{Shafi:2004jy}
K.~S.~Babu, E.~Ma and J.~W.~F.~Valle,
  Phys.\ Lett.\ B {\bf 552} (2003) 207
  [hep-ph/0206292];\\
T.~Kitabayashi and M.~Yasue,
  Phys.\ Rev.\ D {\bf 67} (2003) 015006
  [hep-ph/0209294];\\
  Q.~Shafi and Z.~Tavartkiladze,
  Phys.\ Lett.\ B {\bf 594} (2004) 177
  [hep-ph/0401235];\\
A.~S.~Joshipura,
  hep-ph/0411154;\\
P.~F.~Harrison and W.~G.~Scott,
  Phys.\ Lett.\ B {\bf 594} (2004) 324
  [hep-ph/0403278];\\
N.~Haba and K.~Yoshioka,
  hep-ph/0511108;\\
R.~N.~Mohapatra and W.~Rodejohann,
  Phys.\ Rev.\ D {\bf 72} (2005) 053001
  [hep-ph/0507312]; See also references therein.


\bibitem{Shafi:2005rd}
  Q.~Shafi and Z.~Tavartkiladze,
  Phys.\ Lett.\ B {\bf 633} (2006) 595
  [hep-ph/0509237].



\bibitem{Shafi:2006dm}
  Q.~Shafi and Z.~Tavartkiladze,
  hep-ph/0606188.

\bibitem{Frampton:2002qc}
  P.~H.~Frampton, S.~L.~Glashow and T.~Yanagida,
  Phys.\ Lett.\ B {\bf 548} (2002) 119
  [hep-ph/0208157].



\bibitem{Ibarra:2003xp}
  A.~Ibarra and G.~G.~Ross,
  Phys.\ Lett.\ B {\bf 575} (2003) 279
  [hep-ph/0307051],
Phys.\ Lett.\ B {\bf 591} (2004) 285
  [hep-ph/0312138].


  \bibitem{Guo:2003cc}
  W.~l.~Guo and Z.~z.~Xing,
  Phys.\ Lett.\  B {\bf 583} (2004) 163
  [hep-ph/0310326].

  \bibitem{Guo:2006rm}
  W.~l.~Guo,
  Phys.\ Rev.\  D {\bf 74} (2006) 113009
  [hep-ph/0610174].



\bibitem{Rodejohann:2005ru}
  W.~Rodejohann and M.~A.~Schmidt,
  Phys.\ Atom.\ Nucl.\  {\bf 69} (2006) 1833
  [hep-ph/0507300];\\
  N.~V.~Cortez and M.~D.~Tonasse,
  Phys.\ Rev.\  D {\bf 72} (2005) 073005
  [hep-ph/0510143];\\
  C.~Hagedorn, M.~Lindner and R.~N.~Mohapatra,
  JHEP {\bf 0606} (2006) 042
  [hep-ph/0602244];\\
  Y.~Cai and H.~B.~Yu,
  Phys.\ Rev.\  D {\bf 74} (2006) 115005
  [hep-ph/0608022];\\
  Y.~H.~Ahn, C.~S.~Kim, S.~K.~Kang and J.~Lee,
  Phys.\ Rev.\  D {\bf 75} (2007) 013012
  [hep-ph/0610007];\\
  H.~K.~Dreiner, C.~Luhn, H.~Murayama and M.~Thormeier,
  hep-ph/0610026;\\
  F.~Caravaglios and S.~Morisi,
  hep-ph/0611078.


\bibitem{Albrow:2005kw}
  M.~G.~Albrow {\it et al.},
  hep-ex/0509019.

\bibitem{Alb}
C.~H.~Albright, M.~C.~Chen,
  hep-ph/0608137.


\bibitem{Berezhiani:1996bv}
  Z.~Berezhiani and Z.~Tavartkiladze,
  Phys.\ Lett.\ B {\bf 409} (1997) 220
  [hep-ph/9612232].


\bibitem{Maekawa:2001uk}
  N.~Maekawa,
  Prog.\ Theor.\ Phys.\  {\bf 106} (2001) 401
  [hep-ph/0104200];\\
N.~Maekawa and T.~Yamashita,
  Prog.\ Theor.\ Phys.\  {\bf 108} (2002) 719
  [hep-ph/0205185].



\bibitem{Shafi:1998dv}
  Q.~Shafi and Z.~Tavartkiladze,
  Phys.\ Lett.\ B {\bf 448} (1999) 46
  [hep-ph/9811463].


\bibitem{Shafi:1998jf}
  Q.~Shafi and Z.~Tavartkiladze,
  Nucl.\ Phys.\ B {\bf 552} (1999) 67
  [hep-ph/9807502];
Nucl.\ Phys.\ B {\bf 573} (2000) 40
  [hep-ph/9905202].



\bibitem{Itow:2001ee}
  Y.~Itow {\it et al.},
  hep-ex/0106019;\\
D.~Beavis {\it et al.},
  hep-ex/0205040;\\
D.~Ayres {\it et al.}  [Nova Collaboration],
  hep-ex/0210005.


\bibitem{Borzumati:1986qx}
  F.~Borzumati, A.~Masiero,
  Phys.\ Rev.\ Lett.\  {\bf 57} (1986) 961.


\bibitem{Hisano:1995cp}
  J.~Hisano, T.~Moroi, K.~Tobe, M.~Yamaguchi,
  Phys.\ Rev.\ D {\bf 53} (1996) 2442
  [hep-ph/9510309].

\bibitem{Brooks:1999pu}
  M.~L.~Brooks {\it et al.}  [MEGA Collaboration],
  Phys.\ Rev.\ Lett.\  {\bf 83} (1999) 1521
 [hep-ex/9905013].

\bibitem{Aubert:2005ye}
  B.~Aubert {\it et al.}  [BABAR Collaboration],
  Phys.\ Rev.\ Lett.\  {\bf 95} (2005) 041802
[hep-ex/0502032];
B.~Aubert {\it et al.}  [BABAR Collaboration],
  Phys.\ Rev.\ Lett.\  {\bf 96} (2006) 041801
[hep-ex/0508012].



\bibitem{Fukugita:1986hr}
  M.~Fukugita and T.~Yanagida,
  Phys.\ Lett.\ B {\bf 174} (1986) 45; For non-thermal leptogenesis, see G. Lazerides and Q. Shafi,  Phys.\ Lett.\ B {\bf 258} (1991) 305.


\bibitem{Covi:1996wh}
  L.~Covi, E.~Roulet and F.~Vissani,
  Phys.\ Lett.\ B {\bf 384} (1996) 169
  [hep-ph/9605319];\\
  G.~F.~Giudice, A.~Notari, M.~Raidal, A.~Riotto and A.~Strumia,
  Nucl.\ Phys.\  B {\bf 685} (2004) 89
  [hep-ph/0310123].


\bibitem{Kuzmin:1985mm}
  V.~A.~Kuzmin, V.~A.~Rubakov and M.~E.~Shaposhnikov,
  Phys.\ Lett.\ B {\bf 155} (1985) 36.



\end{thebibliography}

\end{document}